\documentclass[aps,prl,twocolumn,showpacs,amsmath,amssymb,amsfonts,floatfix]{revtex4}

\usepackage[pdftex]{graphicx}
\usepackage{color}
\usepackage{textcomp}

\renewcommand{\section}[1]{\textit{#1} --}

\begin{document}

\title{Half-period Aharonov-Bohm oscillations in disordered rotating optical ring cavities}

\author{Huanan Li, Tsampikos Kottos}
\affiliation{Wesleyan University, Wesleyan Station, Middletown, Connecticut 06459}
\author{Boris Shapiro}
\affiliation{Technion - Israel Institute of Technology, Technion City, Haifa 32000, Israel}

\date{\today}

\begin{abstract}
There exists an analogy between Maxwell equations in a rotating frame and Schr\"odinger equation for a charged particle in the presence 
of a magnetic field. We exploit this analogy to point out that electromagnetic phenomena in the rotating frame, under appropriate conditions, 
can  exhibit periodicity with respect to the angular velocity of rotation. In particular, in disordered ring cavities one finds the optical analog 
of the Al'tshuler-Aronov-Spivak effect well known in mesoscopic physics of disordered metals.
\end{abstract}


\maketitle

{\it Introduction -} When written in the four-dimensional covariant form the Maxwell equations (ME) look the same in any reference 
frame, inertial or not. Things change, however, when the equations are expressed in terms of the three-dimensional vectors designating 
the electric and magnetic fields. For instance, the equation for the electric field in a rotating frame acquires an additional term, as 
compared to its counterpart in an inertial frame (see Eq. (\ref{eq1}) below) \cite{B84}. This additional term associated with a non-zero 
angular velocity of rotation $\Omega\ne 0$, resembles a magnetic flux $\Phi$ experienced by a quantum charged particle when it 
moves through a metallic ring \cite{RR03}. The situation here is similar to that in mechanics, where the equation of motion for a 
particle in a uniformly rotating frame must be supplemented by the Coriolis and centrifugal forces - these forces are often called
``fictitious", although for an observer in the rotating frame they have a very real, observable effect. 

It is therefore natural to expect that the fictitious ``magnetic flux'' term appearing in the framework of electrodynamics of rotating 
media, can lead to important new fundamental insights of light propagation in such set-ups. Of particular interest to us is the case 
of rotating disordered optical ring cavities. The analogous quantum mechanical situation of an electron moving around a ring in the 
presence of magnetic flux results in Aharonov-Bohm (AB) oscillations \cite{AB59} and their half-flux quantum oscillations of conductance 
\cite{AAS81,SS81,CMN84,GIA84,LS86,SI86,AS87}. The latter interference effect occurs only in the presence of disorder and has its 
origin in a coherent backscattering mechanism. These half-flux oscillations have been first predicted by Al'tshuler, Aronov and Spivak 
(AAS) \cite{AAS81} and attracted a lot of theoretical and experimental attention in the frame of mesoscopic quantum physics. A surprising 
finding was that these ``anomalous" oscillations emerge only after an ensemble averaging which restores specific symmetries. 

Do similar effects occur also in the frame of light propagation in rotating disordered ring cavities? As we will show below a direct 
inspection of the ME in the rotating frame uncovers many differences with the quantum mechanical Schr\"odinger equation in the 
presence of a magnetic field. The most noticeable one is that  the fictitious ``magnetic field", related to the rotation, depends not 
only on the angular rotation velocity $\Omega$ but also on the electromagnetic wave frequency $\omega$. We demonstrate that, 
these differences notwithstanding, wave interference effects can lead under appropriate conditions to the optical analog of the AAS 
oscillations. Our theoretical results are supported by detail numerical calculations. 

{\it Model -} We demonstrate the existence of half-period AB oscillations in the presence of rotations using the simple case of 
one-dimensional (1D) ring cavity of radius $R$ \cite{note1}. We assume that the effective refraction index $n(s)$ of the 1D ring generally depends 
on the coordinate $s$ defining the position on the circumference of the ring. The ring rotates uniformly about the axis going 
through its center perpendicular to its plane ($z-$direction), so that the vector angular velocity is ${\vec \Omega}=\Omega 
{\hat z}$. Assuming a monochromatic electric field, having only the $z-$ component, ${\vec E}(s,t)={\hat z} \Psi(s) e^{-i\omega t}$, 
one can write the ME for $\Psi$ {\it in the rotating frame} as
\begin{equation}
\label{eq1}
{\partial^2 \Psi \over \partial s^2} + n^2(s) {\omega^2\over c^2} \Psi - 2i\beta {\omega\over c} {\partial \Psi\over \partial s} = 0
\end{equation}
where $s$ increases from $0$ to $L=2\pi R$ (in the counterclockwise direction) and $\beta\equiv \Omega R/c \ll 1$. This equation 
can be rewritten as
\begin{equation}
\label{eq2}
\left({\partial \over \partial s} -i\beta {\omega\over c}\right)^2 \Psi+ \left[n^2(s)+\beta^2\right] \left({\omega\over c}\right)^2 \Psi = 0
\end{equation}
which makes the analogy with quantum mechanics transparent \cite{note1}. The quantum analogue of the quantity $\beta \omega 
R/c$ is the magnetic flux $\Phi$ through the ring, measured in units of the fundamental quantum flux $\Phi_0=hc/e$. This analogy 
is rather formal because the ``vector potential" in Eq. (\ref{eq2}) depends on the wave frequency $\omega$ which (for the isolated ring) 
is the quantity to be determined from the same Eq. (\ref{eq2}). As a result Eq. (\ref{eq2}) is not periodic with respect to the ``fundamental
flux"-- as opposed to its quantum mechanical equivalent and therefore it is not a-priori obvious that will demonstrate an AB periodicity 
in any observable including its own spectrum. Nevertheless the analogy is useful and it enables us to identify system parameters and 
physical conditions for which AB-type optical interference effects, similar to those known in mesoscopic physics of electronic systems, 
can emerge also in the framework of Eq. (\ref{eq2}). 

Let us consider, for example, a perfect ring cavity with a constant index of refraction $n(s)=n_0$. In the absence of any rotation the spectrum 
is degenerate with eigenfrequencies $\omega_0^{(N)}=Nc/Rn_0$, with $N=1,2,3\dots$. Drawing analogies from the quantum mesoscopic 
physics we expect that significant changes in the spectrum, will occur when the effective magnetic flux $\Phi=\beta \omega_0^{(N)} R/c=
\beta N/n_0$ reaches values comparable to unity. It follows that, due to smallness of $\beta$, this can only happen for large mode 
numbers $N$. For example for a ring of radius $R=3cm$ in the optical range of frequencies $\omega_0^{(N)} \sim 10^{15} sec^{-1}$, 
we have that $N=\omega_0^{(N)}Rn_0/c = 10^{5}$, corresponding to $\Omega\sim 10^5 sec^{-1}$. This number can be decreased 
further to $\Omega \sim 10^3 sec^{-1}$, which is easily achievable, if we consider a loop fiber of length $L\approx 1m$ coiled in 
many turns onto a mandrel.

{\it Isolated ring cavity - } To find the eigenvalues of an isolated ring one has to solve Eq. (\ref{eq2}), with periodic boundary conditions. 
It is instructive to start with an ideal ring, i.e. $n(s)=n_0=$const. For this case the eigenfrequencies, to first order in $\beta$, are 
$\omega_{\pm}^{(N)}=\omega_0^{(N)} \left(1 \mp {\beta\over n_0}\right)$, where the subscript $+ (-)$ denotes a wave propagating 
in (opposite to) the direction of increasing $s$. In the absence of rotation each eigenfrequency is doubly degenerate. Rotations remove 
this degeneracy, introducing frequency splitting $\Delta \omega= 2\beta \omega_0^{(N)}/n_0$. This lifting of degeneracy is similar 
to that in a metallic ring penetrated by flux $\Phi$. There is, however, an important difference: in a metallic ring there is a strict 
universal periodicity, i.e.  energy spectra for $\Phi=0$ and $\Phi=\Phi_0$ are exactly the same. There is no such strict periodicity 
in a rotating optical ring cavity. Indeed, it follows from the above expression for $\omega_{\pm}^{(N)}$ that under increase of $\beta$ 
the doubly degenerate eigenvalue $\omega_0^{(N)}$ is mapped onto $\omega_0^{(N-1)}$ and $\omega_0^{(N+1)}$. The point is 
that this mapping occurs at the value $\beta=n_0/N$ which depends on $N$ itself, so there is no single period for mapping the entire 
spectrum onto itself. 

\begin{figure}[h]
\includegraphics[width=1\columnwidth,keepaspectratio,clip]{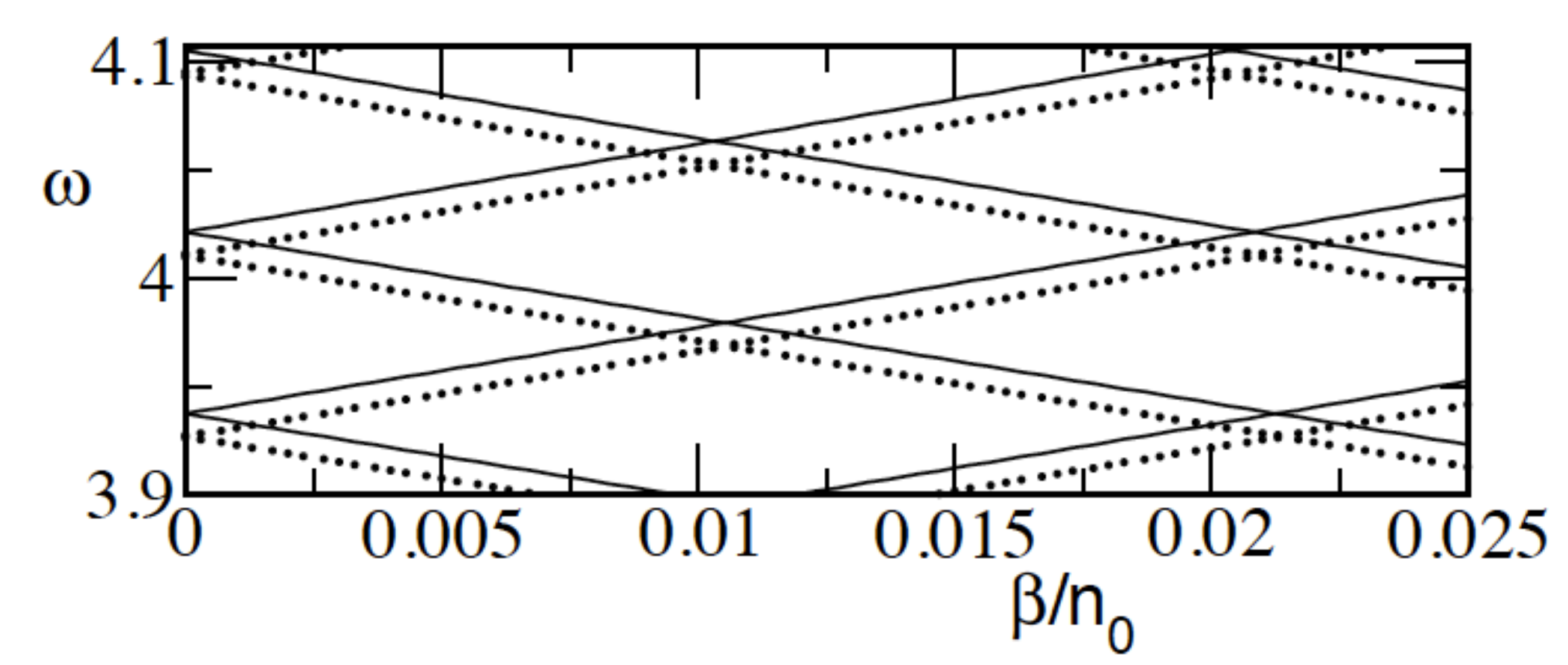}
\caption{Evolution of the spectrum $\omega\left[\frac{c}{l_{1}}\right]$ (we assume $l_{1}=1$) versus $\frac{\beta}{n_{0}}$ for (a) an ideal ring cavity with $n_{0}=1.5$
and circumference $L=50l_{1}$, in the large $N$ limit. Large portion of the spectrum, near $N\approx48$, is accurately mapped onto itself when $\frac{\beta}{n_{0}}=
\frac{1}{N}\approx0.021$. (b) One defect with $n_{1}=1.7$ and length $l_{1}$. In this case the strict degeneracies turn into avoided crossings (dashed lines). 
Note the crossing for $\frac{\beta}{n_{0}}=\frac{1}{2N}\approx0.01$ which play a major role in the disorder-related effect discussed in the text. 
}
\label{fig1}
\end{figure}

The crucial observation is that in the limit of large $N$ one can chose a large set of levels, in an interval $\Delta N$ near $N$, such 
that $1\ll \Delta N \ll N$ for which the period is almost the same, see Fig. \ref{fig1}. Specifically we get a periodicity $n_0/N$ - up 
to small corrections of the order of $\Delta N/N^2$. This period also follows from the aforementioned analogy between $\Phi/\Phi_0$ 
and $\beta \omega R/c$, since $\Phi=\Phi_0$ translates into $\beta = c/R\omega_0^{(N)}=n_0/N$. In addition, a degeneracy at 
$\beta=n_0/2N$ develops when the nearby levels $N$ and $N-1$ cross, see Fig. \ref{fig1}. 

Next we investigate the effect of one defect on the eigenfrequencies $\omega^{(N)}$ of a rotating ring cavity. The impurity splits 
the ring in two sections. The first one consists of the cavity formed by the defect which has an index of refraction $n_1$ and 
length $l_1$. The second one is associated with the rest of the ring and has an index of refraction $n_0$ and length $l_0=L-l_1$ 
($L$ is the length of the ring circumference). We choose the origin $s=0$ in the middle of the section with refraction index 
$n_0$. The interfaces between the two sections are at positions $s_0=l_0/2$ and $s_1=l_0/2+l_1$. In each of the two uniform 
sections, the solution $\Psi_{j}(s)$ of Eq. (\ref{eq1}) can be written as the superposition of two counter-propagating waves,
\begin{equation}
\Psi_{j}(s)=  a_{j}\left(s\right)+b_{j}\left(s\right)
=  a_{j}e^{\imath k_{+}^{j}s}+b_{j}e^{-\imath k_{-}^{j}s},
\label{psi}
\end{equation}
where $k_{\pm}^{j}=n_{j}\frac{\omega}{c}\left(1\pm\frac{\beta}{n_{j}}\right)$ are the wavenumbers associated with a propagating 
wave in (+) or opposite (-) to the direction of increasing $s$ while $j=0,1$ indicates the section associated with the index of 
refraction $n_j$. 

At each position $s$, the field $\Psi(s)$ in Eq. (\ref{psi}) and its derivative are continuous. Implementation of these boundary 
conditions at the interfaces allow us to calculate the transmission $t_+\equiv a_0(L)/a_0(0)$ and reflection $r_+\equiv b_0(0)
/a_0(0)$ amplitudes for a wave traveling in the direction of increasing $s$ and experiencing scattering events 
due to the presence of the defect. Specifically:
\begin{align}
t_{+}= & \frac{4n_{0}n_{1}e^{\imath\frac{\omega}{c}n_{0}l_{0}}}{e^{-\imath\frac{\omega}{c}n_{1}l_{1}}(n_{0}+n_{1})^{2}-
e^{\imath\frac{\omega}{c}n_{1}l_{1}}(n_{0}-n_{1})^{2}}\nonumber \\
r_{+}= & \frac{2\imath\sin\left(\frac{\omega}{c}n_{1}l_{1}\right)\left(n_{1}^{2}-n_{0}^{2}\right)
e^{\imath\frac{\omega}{c}n_{0}l_{0}}}{e^{-\imath\frac{\omega}{c}n_{1}l_{1}}(n_{0}+n_{1})^{2}-
e^{\imath\frac{\omega}{c}n_{1}l_{1}}(n_{0}-n_{1})^{2}}.
\label{onedefect}
\end{align}
Moreover the fields at $s=0$ and at $s=L$ (i.e. across the whole ring) are connected via the total transfer matrix 
$T_{\rm tot}$ \cite{SM08} as
\begin{align}
\begin{pmatrix}
a_0(L)\\
b_0(L)
\end{pmatrix}=T_{\rm tot} 
\begin{pmatrix}
a_0(0)\\
b_0(0)
\end{pmatrix};\,\,
T_{tot} =& \begin{bmatrix}\frac{1}{t_{+}^{*}} & -\frac{r_{+}^{*}}{t_{+}^{*}}\\
-\frac{r_{+}}{t_{+}} & \frac{1}{t_{+}}
\end{bmatrix}e^{\imath\frac{\omega}{c}L\beta},\label{T_tot}
\end{align}
where $t_{+}\equiv\sqrt{T_{+}}e^{\imath\phi_{t}}$ and $r_{+}\equiv \sqrt{1-T_{+}}e^{\imath\phi_{r}}$ can be further parametrized 
in terms of the transmission modulo $T_{+}$ and phase $\phi_t$, and the reflection phase $\phi_r$. 

The eigenmodes and the associated eigenfrequencies are calculated by imposing periodic boundary conditions in Eq. (\ref{T_tot}) 
i.e. $a_0(L)=a_0(0)$ and $b_0(L)=b_0(0)$. The eigenfrequencies of the ring cavity are the solutions of the secular equation 
\begin{equation}
\label{secular}
\det\left(T_{tot}(\omega)-I_{2}\right)=0, 
\end{equation}
where $I_{2}$ is the $2\times 2$ identity matrix. Once the eigenvector $\left(a_0(0),b_0(0)\right)^T$ (corresponding to unity eigenvalue) of 
$T_{\rm tot}$  is evaluated, the associated eigenfunctions are calculated via a straightforward implementation of the 
aforementioned boundary conditions for $\Psi(s)$ and Eq. (\ref{psi}). 

The above formalism can be easily generalized to the case of multiple defects (disordered ring cavity) where now $T_{\rm tot}=
\prod_j T_j$ is a product of transfer matrices $T_j$ \cite{SM08}. The latter takes into consideration the free propagation within 
a uniform defect and the boundary conditions that the wavefunction has to satisfy at the interface between two consequent 
defects. The only $\beta$-dependence in the $T_{\rm tot}$, Eq (5), is in the overall phase, irrespective of the number of defects. 
Therefore the eigenfrequencies (around a constant high $\omega_{\rm avg}$ value) and the corresponding eigenmodes will 
respect a $\beta=2\pi c/\omega_{\rm avg} L$ periodicity.

Substitution of Eq. (\ref{onedefect}) in Eqs. (\ref{T_tot},\ref{secular}) allows us to calculate numerically the eigenfrequencies 
$\omega(\beta)$ of the rotating ring cavity with one defect. The results are shown in Fig. \ref{fig1} together with the eigenfrequencies 
of the uniform ring. We find that the defect lifts the degeneracies at $\beta=0,n_0/2N$ and $n_0/N$. The same
conclusions apply also to disordered rings (not shown). The lift of these degeneracies due to the presence of the defects is crucial 
for the interference effects that we will discuss below.

\begin{figure}[h]
\includegraphics[width=1\columnwidth,keepaspectratio,clip]{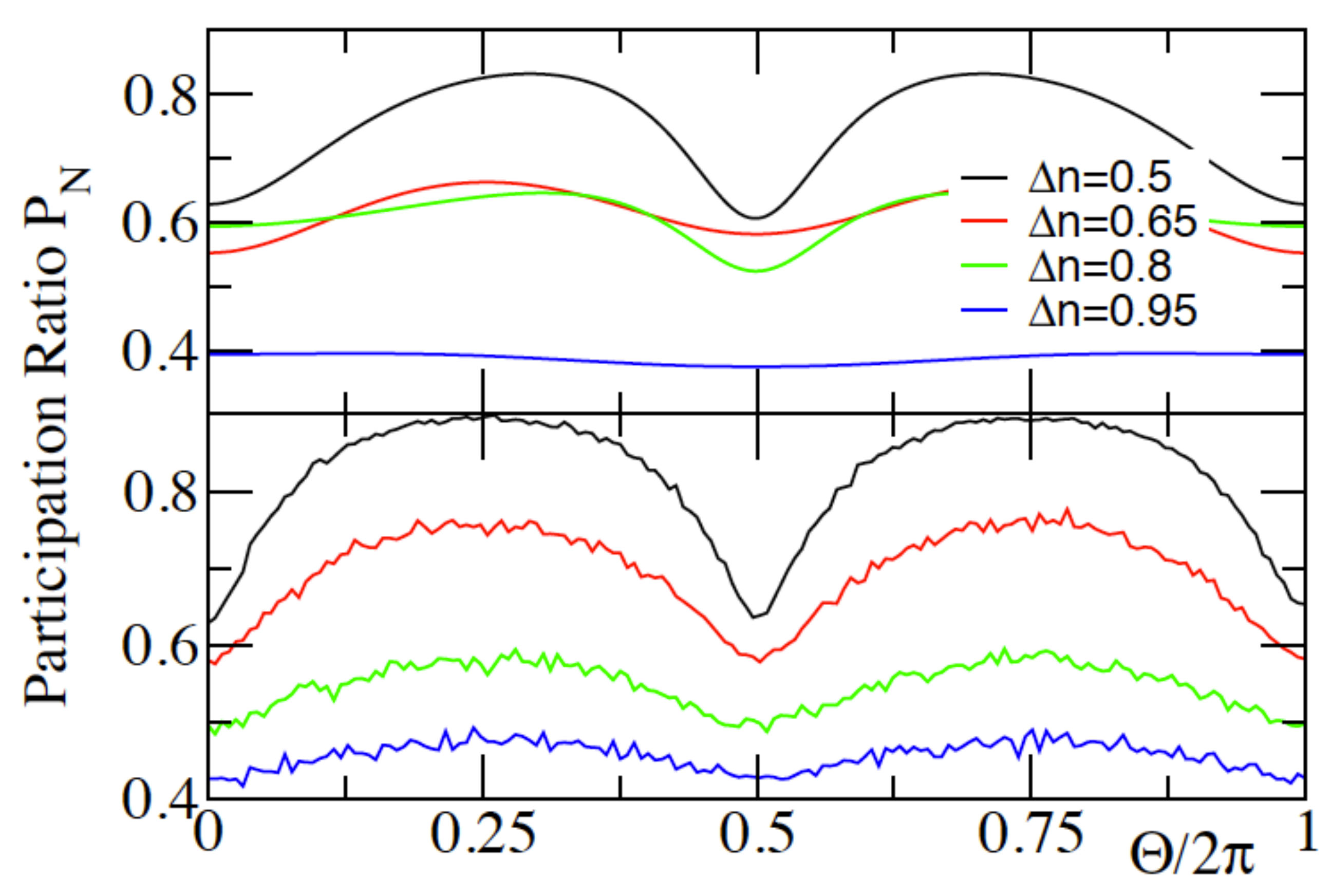}
\caption{Participation ratio $P_{L}$ vs. $\frac{\theta}{2\pi}=\frac{\omega_{avg}}{2\pi c}\beta L$
for a disordered ring with the circumference $L=50l_{1}$. All defects have the same length $l_1$ while their refraction index is given by a box distribution $n\in\left[n_{0}
-\varDelta n,\:n_{0}+\varDelta n\right]$ with $n_{0}=2$. Different colors represent various disordered strengths $\varDelta n$ (indicated in the figure). (a) Single level 
around average angular frequency $\omega_{avg}\approx4\left[\frac{c}{l_{1}}\right]$ for one arbitrary realization. (b) An average over a small 
frequency window around $\omega_{avg}=4\left[\frac{c}{l_{1}}\right]$ and over disordered realizations is taken.
}
\label{fig2}
\end{figure}

Next we turn to the analysis of the eigenmodes and investigate the so-called participation ratio $P_L$
\begin{equation}
P_L(\beta)={1\over L} \langle {\left(\int_0^L ds |\psi^{(N)}(s)|^2\right)^2\over \int_0^L ds |\psi^{(N)}(s)|^4}\rangle
\end{equation}
where $\langle\cdots\rangle$ indicates an average over a small frequency window around a frequency $\omega_{\rm avg}$ 
and over a number of disordered realizations of index of refraction $n(s)\in[n_0-\Delta n, n_0+\Delta n]$. The participation 
ratio allows us to determine the localization of the eigenmodes in the ring and how their spatial distribution within the ring is 
affected by rotations $\beta\ne 0$. Specifically $P_L(\beta)$ takes values between unity (uniformly extended states) and 
zero (exponentially localized states). From the previous discussion it is clear that since the secular equation Eq. (\ref{secular}) 
is periodic with period $\beta=2\pi c/\omega_{\rm avg} L$, the same will apply also for $P_L(\beta)$. This AB-type periodicity 
occurs for {\it each} individual eigenmode, in the absence of any averaging, see Fig. \ref{fig2}a. However, we have found that 
when the ensemble averaging of the participation ratio (for large $\omega$'s) is taken, $P_L(\beta)$ demonstrates a half-flux 
periodicity, similar to the one found in mesoscopic quantum physics \cite{CMN84,AS87}, see Fig. \ref{fig2}b. Further, for strong
disorder, such that $P_L  \ll 1$, the oscillatory behavior of $P_L(\beta)$ completely disappears.

\begin{figure}[h]
\includegraphics[width=1\columnwidth,keepaspectratio,clip]{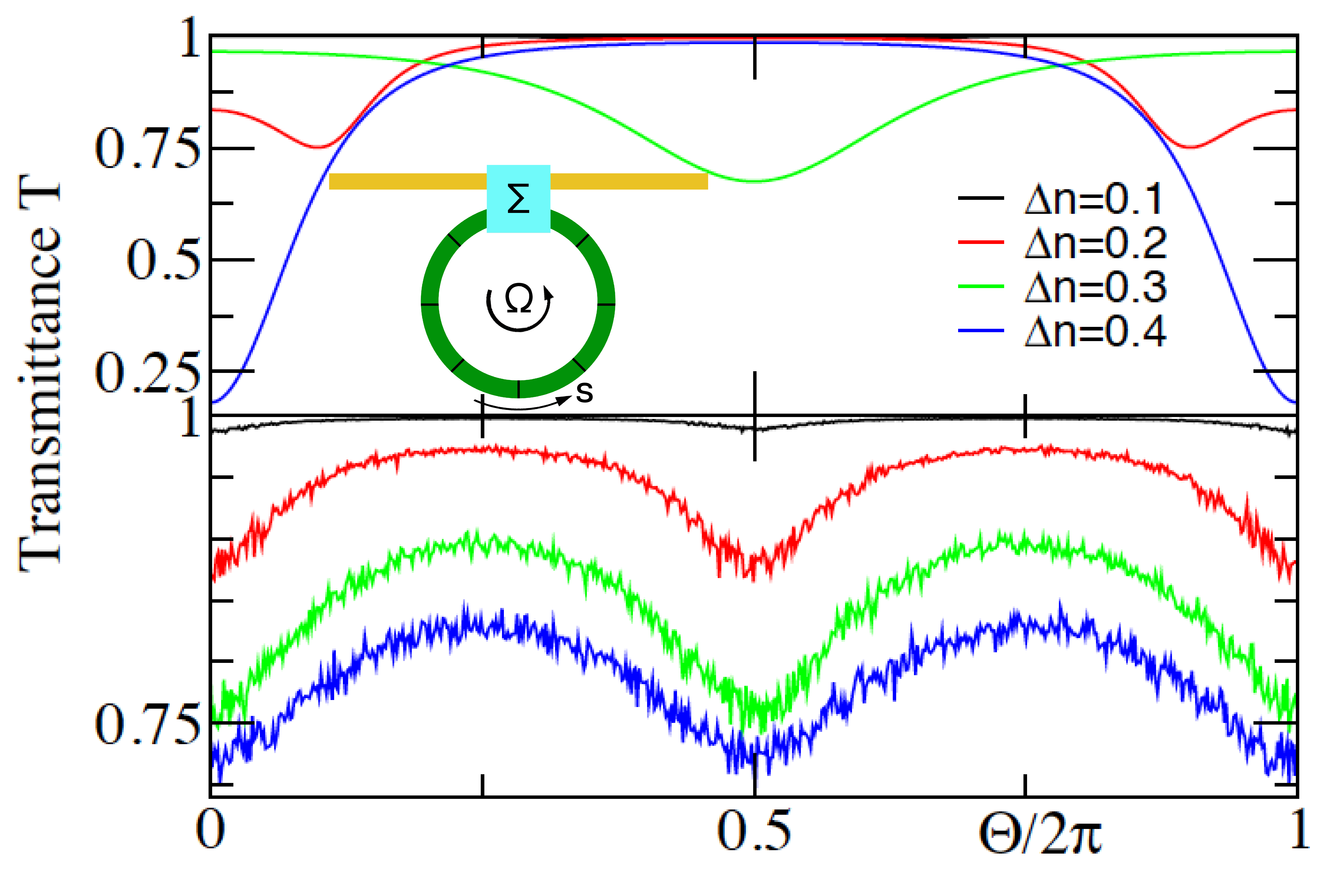}
\caption{Transmittance $T$ vs. $\frac{\theta}{2\pi}=\frac{\omega}{2\pi c}\beta L$ for disordered rings described in Fig. \ref{fig2} with $n_0=1.5$ and $\Delta n$ indicated
in the figure. The frequency of incoming wave is fixed at $\omega=4\left[\frac{c}{l_{1}}\right]$. The coupling parameters between the fiber and the disordered ring are 
$\tau=1/\sqrt{2}$ and $\kappa=1/\sqrt{2}$. Different colors represent various disordered strengths $\varDelta n$ (indicated in the figure). (a) Single realization of disordered 
ring. (Inset: a schematic of the scattering set-up.) (b) An average over $800$ disordered realizations is taken.
}
\label{fig3}
\end{figure}

{\it Scattering set-up -}
 Since properties of the isolated ring eigenmodes are closely related to transmission through the ring, when the latter is connected to external 
leads, we expect oscillating  behavior also in transmittances.
The leads are realized by a straight fiber \cite{note2} coupled with the 1D ring cavity, see the inset of Fig. \ref{fig3}a. 
The coupler between the fiber and the ring cavity is modeled by a $4\times4$ scattering matrix $\Sigma$ defined as \cite{fan}
\begin{align}
\Sigma=\begin{bmatrix}O & K\\
K^{T} & O
\end{bmatrix};\quad
K= \begin{bmatrix}\kappa & \tau\\
-\tau & \kappa
\end{bmatrix},\label{sigma}
\end{align}
where $O$ is the null $2\times 2$ matrix, $\tau\in \left(0,1\right)$, and $\kappa=\sqrt{1-\tau^{2}}$. At the same time the scattering
at the ring is described by the transfer matrix of Eq. (\ref{T_tot}). Combining Eqs. (\ref{T_tot},\ref{sigma}) together leads to the following 
expression for the transmittance $\mathcal{T}$
\begin{widetext}
\begin{eqnarray}
\mathcal{T}= 1-\frac{(1-\tau^{2})^{2}(1-T_{+})}{1+\tau^{4}+2\tau^{2}T_{+}\left(1+\cos\left(2\theta\right)+
\frac{\cos\left(2\phi_{t}\right)}{T_{+}}\right)+4\tau\left(1+\tau^{2}\right)\sqrt{T_{+}}\cos\left(\theta\right)\cos\left(\phi_{t}\right)},
\label{Tringfiber}
\end{eqnarray}
\end{widetext}
where $\theta\equiv\frac{\omega}{c}\beta L$. The above expression applies equally well for both one scatterer and for a disordered ring. 
It follows from Eq. (\ref{Tringfiber}) that $\mathcal{T}\left(\theta+2\pi,\phi_{t}\right)=\mathcal{T}\left(\theta,\phi_{t}\right)$ i.e. for a single 
disorder configuration, the transmittance $\mathcal{T}$ is a periodic function of $\theta$ with period $2\pi$. This
expectation is confirmed by our numerical simulations for single disorder realization, see Fig. \ref{fig3}a. When, however, we 
average ${\mathcal T}$ over disorder realizations we obtain the half period oscillations in the transmittance, in
complete analogy with our previous results for $P_L$. Our numerical results, for various
disorder strengths, are shown in Fig. \ref{fig3}b.

In the weak disorder limit, corresponding to fixed $T_+ (\approx 1)$ and uniform random phase distribution $\phi_t\in [0,2\pi]$, 
one can evaluate the average total transmittance as
\begin{align}
\label{Taver}
\left\langle \mathcal{T}\left(\theta\right)\right\rangle = & \frac{1}{2\pi}\int_{0}^{2\pi}d\phi_{t}\mathcal{T}\left(\theta,
\phi_{t}\right).
\end{align}
Direct substitution of Eq. (\ref{Tringfiber}) in the above equation lead to the AAS relation
\begin{equation}
\langle \mathcal{T}\left(\theta+\pi\right)\rangle =\langle \mathcal{T} \left(\theta\right)\rangle.
\label{Tosc}
\end{equation}
The AAS oscillations are due to interference between a pair of waves traversing the ring in opposite directions and recombining after  
making one (or more) full circle \cite{AS87}. It is important to note that the ``trivial" phase, not related to rotation (i.e. to the ``flux") is exactly the 
same for the two waves and it cancels out in the interference pattern, which is governed by the relative phase equal to $2\theta$. This 
relative phase does not depend on the specific disorder realization which explains the robustness of the ``half-flux" oscillations against 
averaging over disorder. The picture is different for the AB oscillations. For those the topological phase $\theta$ is always accompanied 
by the ``trivial", sample specific phase $\phi_t$. Therefore averaging over disorder leads to decrease of the amplitude of the AB oscillations 
until, for sufficiently strong disorder, they become negligible and only the AAS oscillations remain visible. The above argumentation is
supported by a direct inspection of Eq. (\ref{Tringfiber}) where one sees that the $\cos(2\theta)$-term stands alone, not getting mixed
with terms containing $\phi_t$. In contrast, the $\cos\theta$-term is multiplied by $\cos\phi_t$, and therefore does not survive an
averaging over $\phi_t$.
Finally as the disorder increases further, 
the amplitude of the oscillations decreases and eventually, in the strong localized regime, approaches zero. This is due to the fact that the 
total transmittance in Eq. (\ref{Tringfiber}) involves $T_+$ terms which control the amplitude of the oscillations and which in the 
strong disorder limit go to zero.

{\it Conclusions -} We have shown that the wave equation for light propagating in a rotating ring bears certain analogies with the 
Schr\"odinger equation describing a quantum particle moving in a ring in the presence of a magnetic flux. These analogies have, 
however,  constraints -- the most severe being the fact that the ME is not periodic with the angular velocity, which formally is associated 
with the magnetic flux in the quantum problem. Despite this distinction, we have found that in the optical high frequency domain 
the periodicity of the  spectrum is (approximately) restored. In this frequency domain, AB oscillations emerge for single disorder 
realizations. When a disorder averaging is considered, the fundamental periodicity reduces to half of the AB oscillations. 
Our results are relevant for other classical waves as well. For example the propagation of sound in a randomly corrugated tube 
with a moving fluid is described by similar equations as Eq. (\ref{eq1}) \cite{HATB}. In this framework, AB and AAS oscillations must be 
easily observed for relatively small fluid velocities due to the low values of sound velocities.

{\it Acknowledgements -} This work was partly sponsored by AFOSR MURI Grant No. FA9550-14-1-0037 and by NSF Grant 
No. DMR-1306984.



\begin{thebibliography}{bb}
\bibitem{note1} In the general case where the index of refraction $n(x,y)$ depends on the coordinates $x,y$, the ME for $\Psi$
takes the form $\nabla^2 \Psi + n^2(x,y) {\omega^2\over c^2} \Psi - 2i {\omega\over c^2} ({\vec v}\cdot \nabla)\Psi=0$, where ${\vec v}=
{\vec \Omega}\times {\vec r}$. This equation can be re-written in a form similar to Eq. (\ref{eq2}) i.e. $\left(\vec{\nabla}-i{\omega\over c}
{{\vec \Omega}\times {\vec r}\over c}\right)^2\Psi+\left[n^2(x,y)+\left({{\vec \Omega}\times {\vec r}\over c}\right)^2\right]{\omega^2\over 
c^2}\Psi=0$. One can make use of ``gauge freedom" in the last equation, in order to simplify treatment in certain geometries. For instance, 
if $n(x,y)$ depends only on one coordinate, then it is clearly advantageous to use Landau gauge rather than circular gauge. This gauge 
freedom also enables one to construct solutions of the ``clean" problem (i.e. for $n(x,y)=n_0$=const.) which can be useful in defining the 
appropriate asymptotic states in a scattering set-up.
\bibitem{note2} A wave propagating at the distance $s=x_1-x_0$ along the fiber is accumulating an additional phase 
$\exp\left(\imath\frac{\omega}{c}\int_{x_{0}}^{x_{1}}dx\frac{\hat{x}\cdot\left(\overrightarrow{\varOmega}\times\overrightarrow{r}\left(x\right)\right)}{c}\right)$ 
due to the rotation, where $\hat{x}$ denotes the
unit vector along the fiber and $\overrightarrow{r}\left(x\right)$
is the position vector of position $x$ in the fiber with respect
to the rotation center. This phase does not affect the calculation of the transmittance.
\end{thebibliography}
\end{document}